\documentclass[aps,twocolumn,amsmath,amssymb,floatfix,nofootinbib]{revtex4-2}
\usepackage{graphicx}
\usepackage{placeins}
\usepackage{dcolumn}
\usepackage{bm}
\usepackage{mciteplus}
\usepackage{braket}
\usepackage{upgreek}
\usepackage{amsmath}
\usepackage{amssymb}
\usepackage{natbib}
\usepackage[svgnames]{xcolor}

\usepackage[colorlinks=true,linkcolor=DarkBlue,urlcolor=DarkBlue,citecolor=FireBrick]{hyperref}

\begin{document}
	\title{Description of laminar-turbulent transition of an airfoil boundary layer \\measured by differential image thermography using directed percolation theory}
	
	\author{T.~T.~B.~Wester$^1$}
	\email{e-mail: tom.wester@uni-oldenburg.de}
	\author{J.~Peinke$^1$} 
	\author{G.~G\"{u}lker$^1$}
	\affiliation{$^1$ForWind, Institute of Physics, University of Oldenburg, Oldenburg, Germany}
	
	\date{\today}
	
	\begin{abstract}
		The study presented here addresses the challenging problem of laminar-turbulent flow transition in boundary layers. Directed percolation theory has emerged as a promising approach to understand and describe this transition in different scenarios. This study utilizes differential image thermography (DIT) to investigate the boundary layer transition on the suction side of a heated airfoil, presenting new experimental findings. First, the DIT results underline the ability of capturing the near surface transition for the airfoil boundary layer with a high temporal and spatial resolution. Second, the evaluation reveals the effectiveness of directed percolation theory in describing the onset of the transition, showing agreement with all three universal exponents of (1+1)D directed percolation theory. Third, the study shows the applicability of this theory to a wide range of flow situations beyond the parameter space covered in previous examinations. These findings underscore the possible application of directed percolation models in fluid mechanics and suggest that the theory could serve as a high precision tool for describing the transition to turbulence.
	\end{abstract}
	\maketitle	
	
	\section{Introduction}
	% Transition as a general problem in turbulence
	%
	Laminar-turbulent transition is a complex phenomenon that has been one of the most challenging problems in turbulence research since its first observation in Reynolds' famous pipe flow experiment in 1883 \cite{Reynolds1883}. Despite considerable efforts in the field, the standard stability theory has failed to explain the occurrence and nature of transition even for basic cases. Reviews on the ongoing research on wall-bound transition have been presented by Manneville \cite{Manneville2016} and Barkley \cite{Barkley2016}. \\	
	% Idea of Pomeau
	%
	To simplify the complexity of the problem, Pomeau proposed a novel approach to describe the non-linear transition phenomena of a flow as interacting oscillators more than three decades ago \cite{pomeau1986front}. Pomeau's approach is based on subcritical bifurcation of the flow in the region of the critical Reynolds number, where the flow switches between a stable and a meta-stable state. This new perspective on the transition has started discussions on whether the laminar-turbulent transition fits into one of the universality classes of the statistical model of directed percolation.\\
	Pomeau's idea is appealing since the laminar-turbulent transition can be described as a spatio-temporal intermittency \cite{chate1987transition,rupp2003critical}. However, it took several decades before percolation theory was taken up again in fluid dynamics, although the model has been successfully applied to other fields such as epidemics and forest fires \cite{hinrichsen2000non}. Pomeau attributed this delay to the lack of accuracy of experimental and numerical studies in fluid dynamics available at that time, making it impossible to verify the theory \cite{pomeau2016long}. Therefore, high-resolution spatio-temporal methods are required for laminar-turbulent transition experiments, while numerical flow simulations must be high resolving and sufficiently long for conclusive statistical evaluations.\\	
	% First measurements using the Percolation theory 
	%
	\footnote{For an explanation of the following terminology used here, please refer to the later chapter \ref{chap:Percolation} or for more general information to \cite{Grimmett_1999,hinrichsen2000non}.}In recent years, the availability of better numerical and experimental resources has led to an increase in the number of studies focusing on directed percolation theory in fluid dynamics. These studies have investigated the influence of the order parameter on the propagation of turbulent cells \cite{allhoff2012directed,sipos2011directed}, the formation and development of clusters of turbulent cells \cite{kreilos2016bypass}, and the determination of turbulent cells from simulations \cite{rehill2013identifying}. In addition, low-order models for describing the transition have been developed \cite{barkley2011simplifying}, sophisticated simulations have been performed \cite{shih2016ecological,avila2013nature}, and the applicability of percolation theory has been demonstrated in a broad range of experiments \cite{barkley2015rise,lemoult2016directed,Sano2016,wester2017percolation,traphan2018aerodynamics}. More recent studies have focused on perturbing coherent structures and their influence on percolation results \cite{manneville2019subcritical}. This accelerating interest in the field is also reflected in various conference contributions \cite{lemoult2020experimental,shih2020directed} and publications \cite{Klotz_2022} in the recent years. The progress is summarized in the review of Hof \cite{Hof_2022}.\\
	% Airfoil experiment PRX 
	%
	The majority of studies in this field focus on canonical wall-bounded flows with basic geometries, such as Couette or pipe flows, where changes to the Reynolds number of the entire system are achieved by altering velocities or aspect ratios. Another application of the theory has been found for the transition of an airfoil boundary layer flow into the onset of a laminar separation bubble using stereoscopic Particle Image Velocimetry (PIV) \cite{traphan2018aerodynamics}.
	Unlike other studies, the flow evolves in an instationary way along the surface of the airfoil. Thus a changing Reynolds number based on the traveled distance was used to characterize the system, which is often done for developing boundary layer flows. The data showed good agreement with the (1+1)D directed percolation class, where one dimension describes the spatial component, and the other the temporal evolution. The study presented results for one specific angle of attack and inflow velocity. Additionally, since the laminar separation bubble case is a unique instance, it would be interesting to find out whether the more frequent natural laminar-turbulent transition of the airfoil boundary layer can also be described by a percolation universality class. The applicability of the method in the context of experimentally variable parameters and its potential deviations from the universality of directed percolation, which should not occur, are still open questions.\\
	% Importance of airfoil measurements and understanding due to application
	% 
	In general, an accurate determination and description of boundary layer transition and its location holds great importance in fluid mechanics applications for the development of more accurate models. The use of a simple statistical model like directed percolation can provide a more precise description, thereby improving the prediction of loads on rotor blades and enhancing the reliability of CFD simulations of airfoil behavior. Underestimated loads often arise from missing the increased energy input of turbulent boundary layer into the near-surface flow, resulting in a longer-attached flow and delayed and stronger flow separation. Similarly, transition can lead to a sudden increase in loads when an already detached flow suddenly reattaches. These rapidly changing forces are undesirable in most applications and must be understood and included into prediction models. Therefore, directed percolation theory has the potential to become a valuable tool in predicting, locating, understanding, and modeling the transition phenomenon in the future.\\	
	% Challenges in measuring the airfoil boundary layer (curved surface etc.)
	%
	To determine whether directed percolation theory can accurately describe the laminar-turbulent transition of a boundary layer, it is essential to ensure a high level of temporal and spatial resolution during experiments. However, this presents a significant challenge when dealing with airfoils, as their curved surfaces and three-dimensional boundary layers complicate the measurement process. Planar PIV measurements utilizing light sheets can only provide limited insights, as the distance between the curved surface and measurement volume varies, leading to inconsistent measurements of different boundary layer heights and states. This issue has been identified as a potential cause of deviations in previous studies \cite{traphan2018aerodynamics}, and must be overcome in order to achieve precise analysis.\\
	% Using differential image thermography to solve the problem
	%
	Differential image thermography (DIT) has proven itself to be an effective tool to investigate complex boundary layer flows. Numerous studies have demonstrated the potential of DIT in accurately distinguishing between laminar and turbulent flows in unsteady boundary layers \cite{DeLuca1990,raffel2014differential,gardner2017analysis,wolf2019optimization}. This method has been successfully employed to investigate the boundary layer flow over pitching airfoils \cite{raffel2015differential} as well as the phenomenon of dynamic stall \cite{gardner2016new}. Furthermore, DIT has been applied to examine boundary layer transition of rotating rotor blades, at both small \cite{raffel2017rotating} and large scales \cite{reichstein2019investigation,dollinger2018measurement}. With its diverse applications and the growing capabilities for temporal and temperature resolution, the DIT has demonstrated its versatility and capacity to offer in-depth insights into complex flow situations, all while maintaining an easy operation. As a result, this technique has become increasingly promising for both qualitative and quantitative analyses, such as directed percolation.\\
	% Present Study 
	% 
	The objective of this study is to measure the laminar-turbulent transition of an airfoil boundary layer at different inflow velocities and angles of attack to cover a broad parameter space. To overcome the challenge of the curved surface and obtain detailed insight into the boundary layer flow, the study utilizes DIT. By measuring surface temperature differences, the location of the transition can be determined. The study demonstrates that DIT offers adequate spatio-temporal resolution for conducting statistical evaluations and, ultimately, enables the comparison between (1+1)D directed percolation model and boundary layer evolution. The results demonstrate a very good agreement between theory and transition for a broad range of parameters, leading to a more precise localization of the transition point compared to state of the art approaches.\\ 
	% Structure
	%
	The paper starts by the description of  the experimental setup in Sec. \ref{chap:Setup}, which comprises the wind tunnel and the thermography system. The DIT results are then presented in chapter Sec. \ref{chap:Thermography}, followed by the application of the directed percolation theory to the obtained data in chapter Sec. \ref{chap:Percolation}. Sec. \ref{chap:Summary} summarizes the paper.	
	
	\section{\label{chap:Setup}Experimental Setup}	
\begin{figure*}[htbp]
		\begin{centering}
			\includegraphics[width=6in]{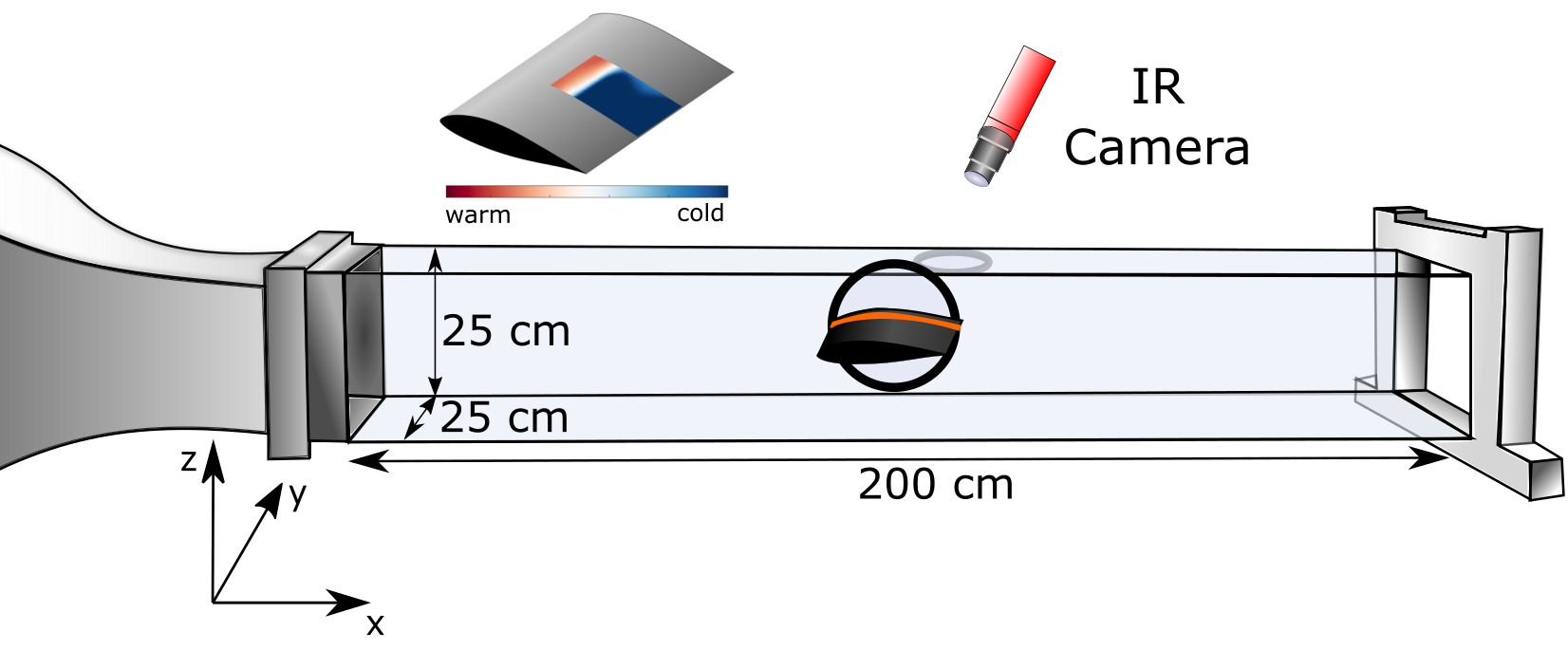}
			\caption{Experimental setup consisting of a Göttingen type wind tunnel, thermography camera shown in red, and a DU95-W-180 airfoil mounted on a rotary table. Above the test section, the airfoil with the FOV of the thermography camera is represented by an example differential image thermogram.}
			\label{fig:1}
		\end{centering}
	\end{figure*}

	% wind tunnel
	An experimental setup utilizing a Göttingen-type return wind tunnel located at the University of Oldenburg is used, as illustrated in Fig. \ref{fig:1}. The wind tunnel is equipped with a PID velocity control system to maintain the desired flow conditions. The test section of the wind tunnel is enclosed and made of acrylic glass for optical access. The test section has dimensions of 0.25m $\times$ 0.25m $\times$ 2.00m (height $\times$ width $\times$ length). To enable measurements in the infrared wavelength range using the differential infrared thermography (DIT) technique, a calcium fluoride window with infrared transmission greater than 90\% is embedded in the ceiling of the test section. Its position can be adjusted to optimize the camera's angle of view on the object being studied.
	 \\
	A DU95-W-180 airfoil with a chord length of 180mm and made of aluminum for high heat capacity is selected for the experiments. The airfoil is heated from the inside using a hot air flow to achieve a temperature difference of about 5K to the equilibrium state, which provides a sufficiently large contrast during the measurement without fundamentally changing the aerodynamics on the boundary layer of the profile. A black foil is applied to the airfoil's surface to avoid reflections of the aluminum surface.\\
	A turntable is embedded in the sidewall of the test section 1m downstream of the nozzle. This allows the mechanical change of the angle of attack ($\alpha$) of the airfoil. The surface temperature is measured by a thermography camera of type ImageIR 8380 hpS by Infratec. The settings of the camera enable an accuracy of 20mK with a temporal resolution of up to 355fps using the entire sensor chip with a size of 640px $\times$ 512px. Each data set captured consists of 1775 images or a total duration of 5s. The field of view (FOV) has a dimension of 107.9mm $\times$ 49.8mm with a spatial resolution of 0.172mm/px (e.g. $\alpha$=0$^\circ$). The captured FOV is illustrated on the airfoil surface in Fig. \ref{fig:1}. To calculate differential image thermograms, a image of the heated airfoil is taken for each measurement series before the wind tunnel is switched on. This thermogram serves as a reference image for the subsequent determination of the temperature loss induced by the developing boundary layer flow. In general, the airfoil experiences cooling as a consequence of the airflow, leading to the development of a discernible temperature over the measurement time. In order to mitigate the potential influence of this on the research outcomes, the measured data are subjected to a correction procedure involving the removal of the linear trend associated with the temporal temperature evolution. This corrective methodology ensures that the obtained results are unaffected by the presence of an overall heating or cooling of the system, thereby enhancing the accuracy and reliability of the findings.\\
	The study investigates a range of boundary layer states by varying the velocity from $u_\infty=$ 10m/s to $u_\infty=$ 20m/s and setting $\alpha$ of the airfoil to $[0,5,10]^\circ$ successively. This allows for the observation of not only transition but also the onset of static stall during the investigation.\\
	
	\section{\label{chap:Thermography} Thermography measurements}
\begin{figure*}[htbp]
		\centering
		\includegraphics[width=6in]{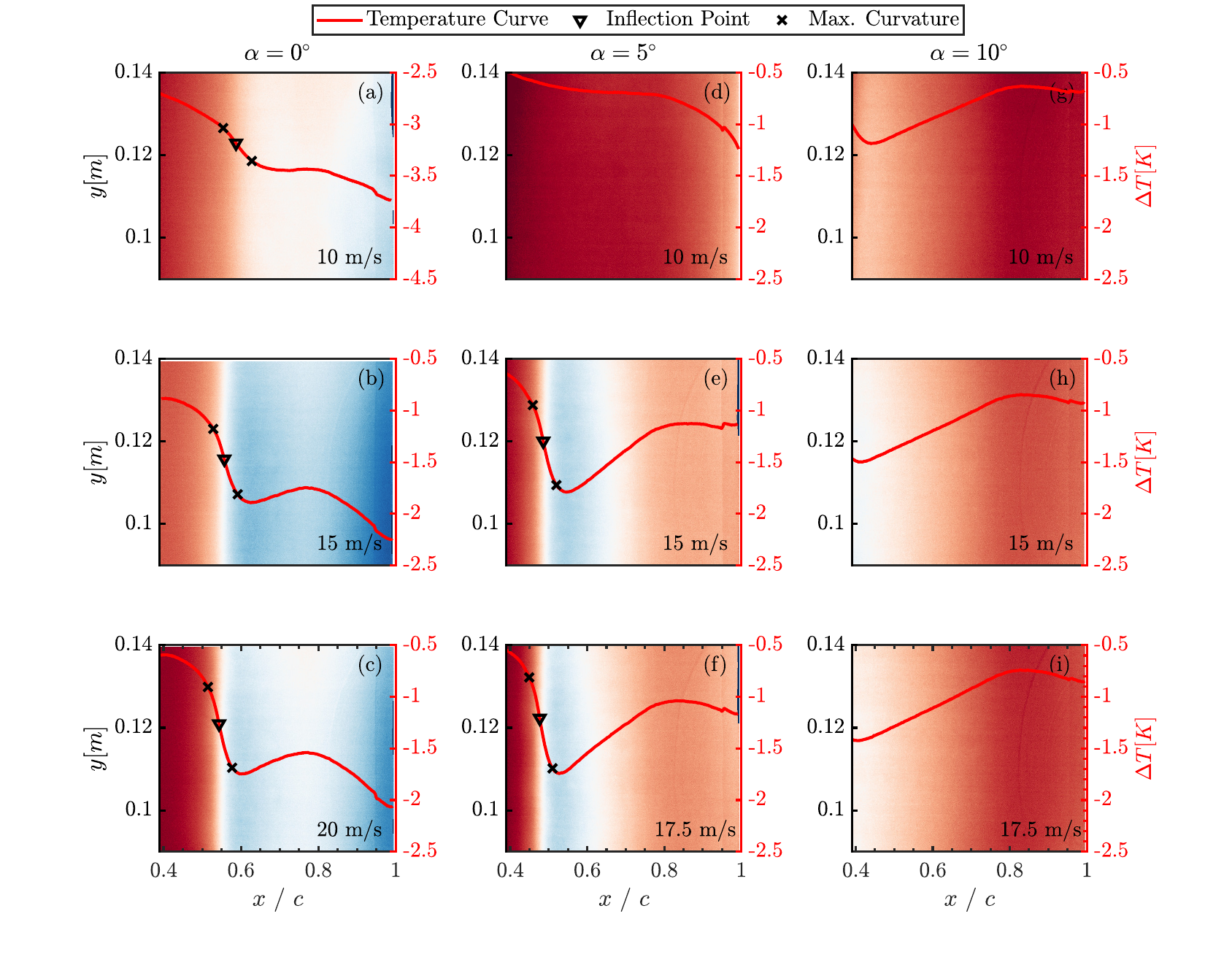}
		\caption{Time averaged differential image thermograms for different angle of attack $\alpha=[0, 5, 10]^\circ$ and velocities $u_\infty = [10,15,17.5,20]m/s$. The colors represent the temperature from cold (blue) to warm (red). The right $y$-axis of each plot corresponds to the color bar range. Additionally, the spatial averaged temperature curves are plotted in red. The black symbols represent the inflection point $\triangledown$ and the point of maximum curvature $\times$.}
		\label{fig:3}       % Give a unique label
	\end{figure*}

	Boundary layer flows are analyzed using DIT by exploiting the different heat transfer properties between the object surface and the fluid flow, which are characterized by the Stanton number
	\begin{equation}
		St := \frac{h}{u_\infty \ \rho_{\rm{Fluid}} \ c_P},
	\end{equation}
	where $h$ is the convection heat transfer coefficient, $u_\infty$ is the flow velocity, $\rho_{\rm{Fluid}}$ is the fluid density, and $c_P$ is the fluid specific heat.\\
	In laminar flows, $h$ and thus $St$ are smaller compared to turbulent flows, due to the slower dissipation of heat \cite{raffel2015differential}. This leads to increased cooling rates of surfaces for turbulent boundary layers. Instead of measuring $h$ or $St$, the inverse proportional and easy accessible temperature $T$ can be measured. By this the evolving temperature development can be used to directly infer the flow state of the boundary layer.\\
	To determine the transition point $\widetilde{Re_c}$ from DIT measurements, the temperature evolution along the suction side of the airfoil is examined. Fig. \ref{fig:3} shows a color-coded representation of the time-averaged differential temperatures $\Delta T$. The differential images are calculated by subtracting the beforehand measured equilibrium state from each thermogram. Each column represents a different angle of attack ($\alpha$), and the each row represents different used wind speeds ($u_{\infty}$). The red line shows the temperature gradient averaged along the span wise $y$-direction of the airfoil.\\
	Based on the measured data, it is observed that the DITs can be classified into two distinct groups. The first group shows a temperature drop and a local minimum along the chord (as depicted in Fig. \ref{fig:3} (a), (b), (c), (e), (f)), while the second group either displays a slow temperature decrease (as shown in Fig. \ref{fig:3} (d)) or a monotonic increase (as illustrated in Fig. \ref{fig:3} (g), (h), (i)). The temperature gradients observed in the first group are characteristic of laminar-to-turbulent boundary layer transitions \cite{DeLuca1990}. During the initial laminar flow regime, the associated low Stanton number $St$ results in less cooling of the surface. However, as the flow transitions to turbulence, $St$ increases. Due to the increased heat transfer coefficient $h$, the surface temperature decreases faster. Using this information, the measured temperature profiles can be utilized to locate the point of transition.
	Such locations on the airfoil are given in terms of chord length based Reynolds number 
	\begin{equation}
		Re_x := \frac{u_\infty x}{\nu},
		\label{eqn:1}
	\end{equation}
	where $x$ denotes the position along the chord, $u_\infty$ is the free stream velocity and $\nu$ represents the kinematic viscosity of used fluid.\\
	In accordance with contemporary best practices to find the transition point \cite{DeLuca1990,raffel2015differential}, inflection points for each temperature profile are determined initially by calculating the minimum of the first derivative, represented as $min\left(\frac{d \Delta T}{dx}\right)$. The positions of these inflection points are indicated using diamond symbols in Fig. \ref{fig:3}. Given that the temperature transition in this context is not characterized as an abrupt step but rather as a gradual decay, points of maximum curvature are also derived from the second derivative, specifically $min\left(\frac{d^2 \Delta T}{dx^2}\right)$ and $max\left(\frac{d^2 \Delta T}{dx^2}\right)$, based on the displayed temperature profiles. These points serve to estimate the region where the transition occurs and are denoted by crosses in Fig. \ref{fig:3}. The values of the transition points ($\widetilde{Re_c}$) obtained from DIT images are summarized in Tab. \ref{tab:1}, with the errors $\Delta \widetilde{Re_c}$ derived as the minima between the inflection points and the points of maximal curvature. \\
	Next, the two previously defined groups are compared in further detail. In contrast to the findings of $\alpha = 0^\circ$, the measurements show at $\alpha = 5^\circ$ and at $x/c \approx 0.7$ a very pronounced rise in temperature  (refer to Fig. \ref{fig:3} (e), (f)). This finding implies the existence of an additional flow phenomenon, apart from the laminar-to-turbulent transition. The temperature increase can be ascribed to a decrease in $St$ resulting in reduced heat transfer. One plausible explanation for this phenomenon is the formation of a stall cell at the trailing edge of the airfoil, which leads to a diminished surface velocity. This explanation aligns with previous studies on airfoil characteristics available in the literature \cite{timmer2003summary}.\\	
	In the case of the second group of measurements, including measurement of Fig. \ref{fig:3}(d), no decrease in temperature is observed. The temperature profiles indicate that the transition from laminar to turbulent flow had already occurred upstream of the field of view (FOV) investigated by the DIT. Consequently, the precise location of the transition point cannot be determined using the available DIT data and the curvature method. Nevertheless the temperature measurements for $\alpha =10^\circ$  reveal a uniform increase in the temperature gradient. This observation suggests the presence of a stall phenomenon at this angle, which aligns with results from previous investigations \cite{timmer2003summary}.
	\begin{table}[htbp]
		\centering
		\caption{Transition point $\widetilde{Re_c}$ extracted from DIT measurements referring to Fig. \ref{fig:3} and transition points $Re_c$ extracted from (1+1)D directed percolation model fit shown in Fig. \ref{fig:7}.}
		\label{tab:1}% 
		\begin{tabular}{|c|c|c|c||c|c|c|}
			\hline
			& & &   & & &\\			
			& $\widetilde{Re_c}$ & $\Delta \widetilde{Re_c}$ & $\Delta \widetilde{Re_c}/\widetilde{Re_c}$ &$Re_c$ & $\Delta Re_c$ & $\Delta Re_c/Re_c$  \\
			\hline
			(a) & 66909  & 3770 & 5.6\% &66809 & 43 & 0.06\%\\
			\hline
			(b)& 95216 & 4903 & 5.1\%  &94976 & 175 & 0.18\%\\
			\hline
			(c)& 123845 & 6569 & 5.3\%  & 123720 & 125 & 0.10\%\\
			\hline
			(d)& $<$44049 & - & -  &43716 & 92 & 0.21\%\\
			\hline 
			(e)& 83043 & 4658 & 5.6\%  &83020 & 92 & 0.11\%\\
			\hline
			(f)& 95057 & 5449 & 5.7\%  &94999 & 200 & 0.20\%\\
			\hline
			(g)& $<$ 42151& - & -  &42009 & 17 & 0.04\%\\
			\hline
			(h)& $<$ 62946& - & -    &62681 & 63 & 0.10\%\\
			\hline
			(i)& $<$71747 & - & -    &71489 & 96 & 0.13\%\\
			\hline
		\end{tabular}
	\end{table}%
	
	\section{\label{chap:Percolation} Measured laminar-turbulent transition in terms of percolation theory}
\begin{figure}[h!]
		\centering
		\includegraphics[width=3.1in]{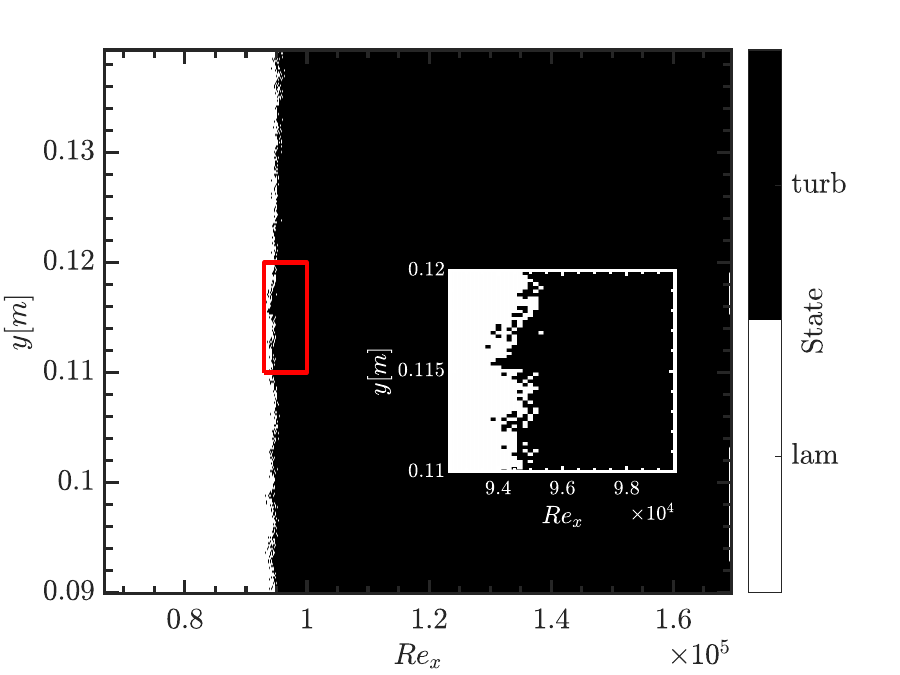}
		\caption{Binarized differential image thermogram of measurement Fig. \ref{fig:3} (b) using the defined threshold. Inlay shows a zoom into the transition area marked by the red box.}
		\label{fig:5} 
	\end{figure}

\begin{figure}[h]
		\includegraphics[width=3.1in]{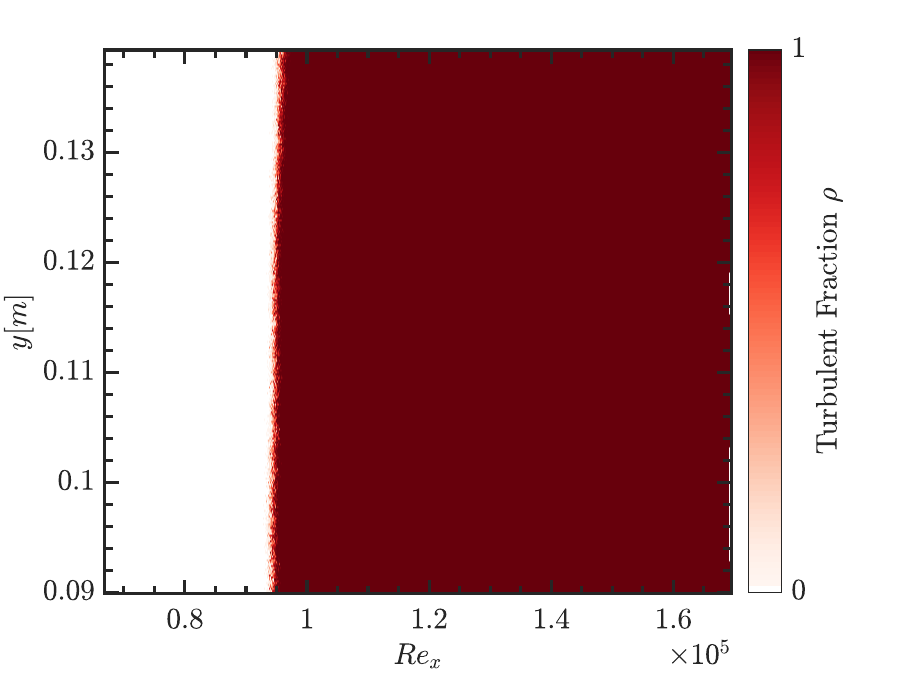}
		\caption{Turbulent fraction $\rho$ calculated from the binarized DITs of measurement Fig. \ref{fig:3} (b).}
		\label{fig:6}
	\end{figure}

\begin{figure}[h]
		\centering
		\includegraphics[width=3.1in]{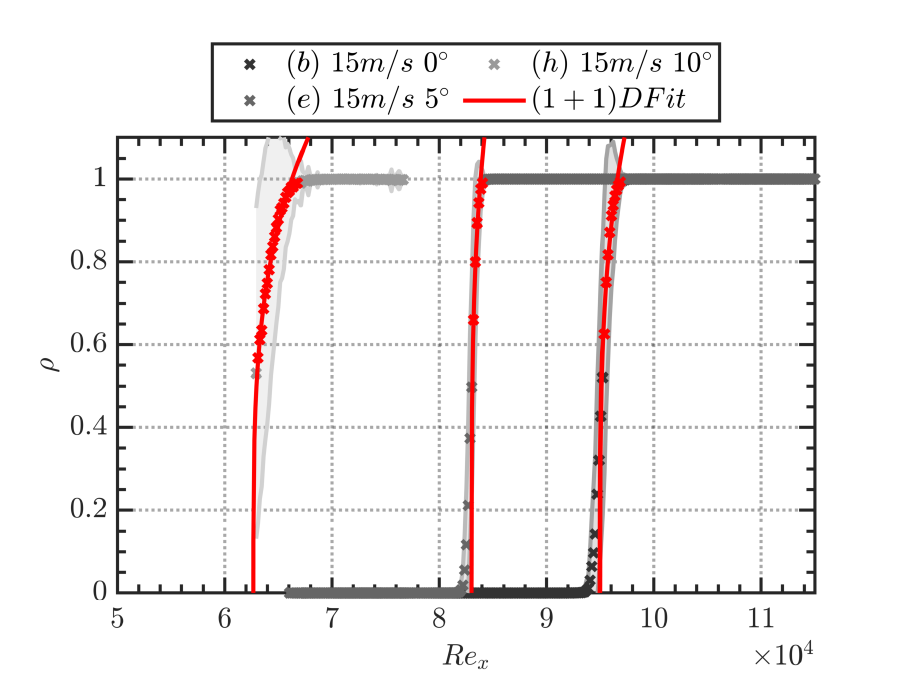}
		\caption{Turbulent fraction $\rho$ as result of taking the average along the y-direction from Fig. \ref{fig:6}. Shown error bars represent the standard deviation of the averaging. Red curves show a fit using Eqn. (\ref{eqn:2}). The red symbols on each curve represent the points used for fitting.}
		\label{fig:7}
	\end{figure}

\begin{figure}[h!]
		\centering
		\includegraphics[width=3.1in]{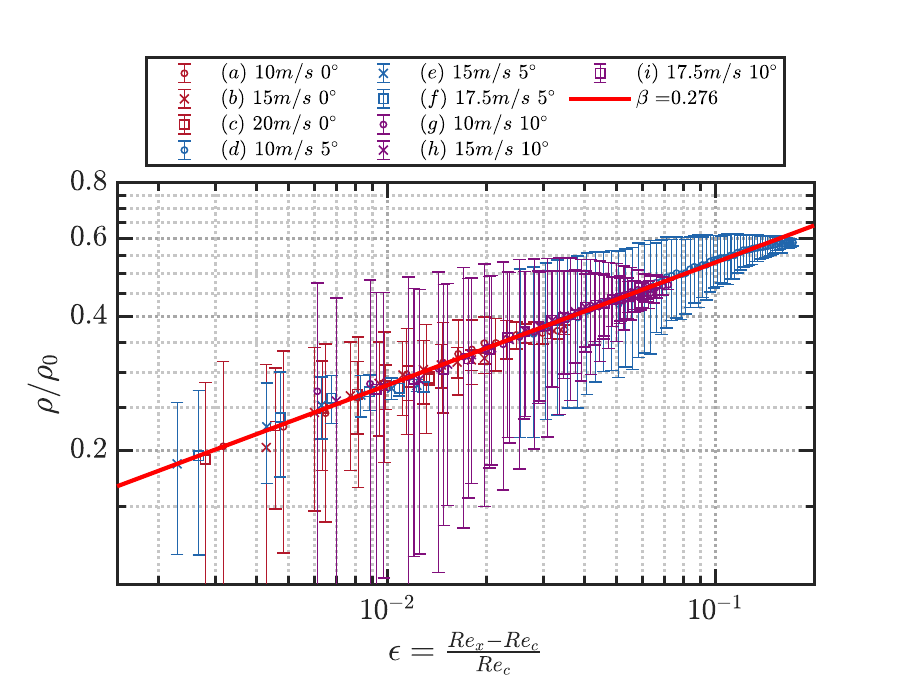}
		\caption{Logarithmic representation of the normalized turbulent fraction $\rho$ over $\epsilon= \frac{Re_x-Re_c}{Re_c}$ for all performed measurements. The red solid line represents the theoretical slope of $\beta$.}
		\label{fig:LogLogTurbFrac}
	\end{figure}

\begin{figure}[h!]
		\centering
		\includegraphics[width=3.1in]{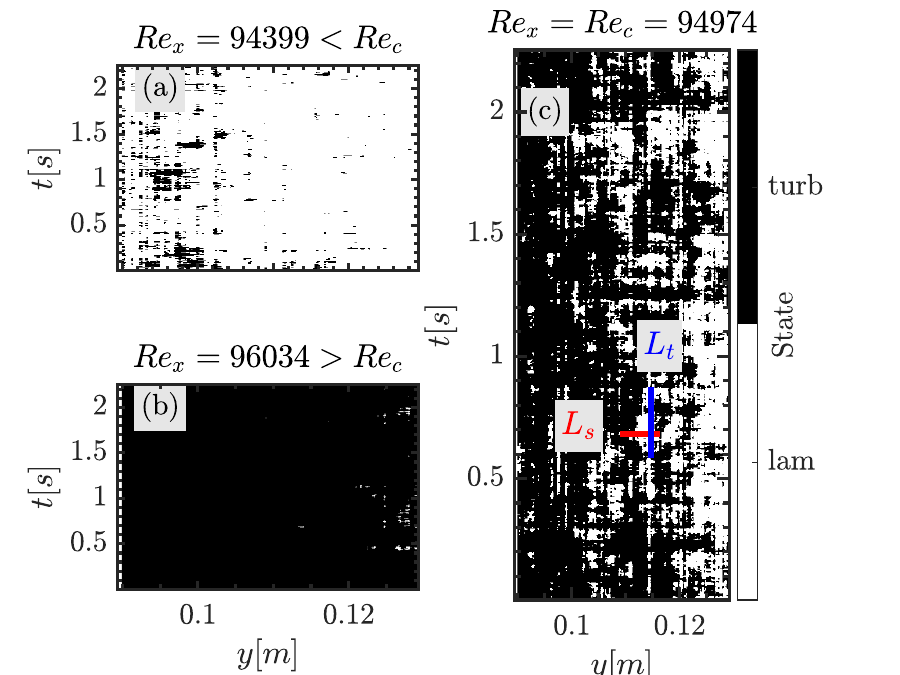}
		\caption{(a) Field for $Re_x < Re_c$. (b) $Re_x > Re_c$ and (c) $Re_x = Re_c$. (c) also includes an exemplary representation of the definition of cluster sizes in spatial ($L_s$) and temporal ($L_t$) direction. This figure is based on measurement Fig. \ref{fig:3} (b).}
		\label{fig:Cluster_extraction}
	\end{figure}

\begin{figure}[h!]
		\centering
		\includegraphics[width=3.1in]{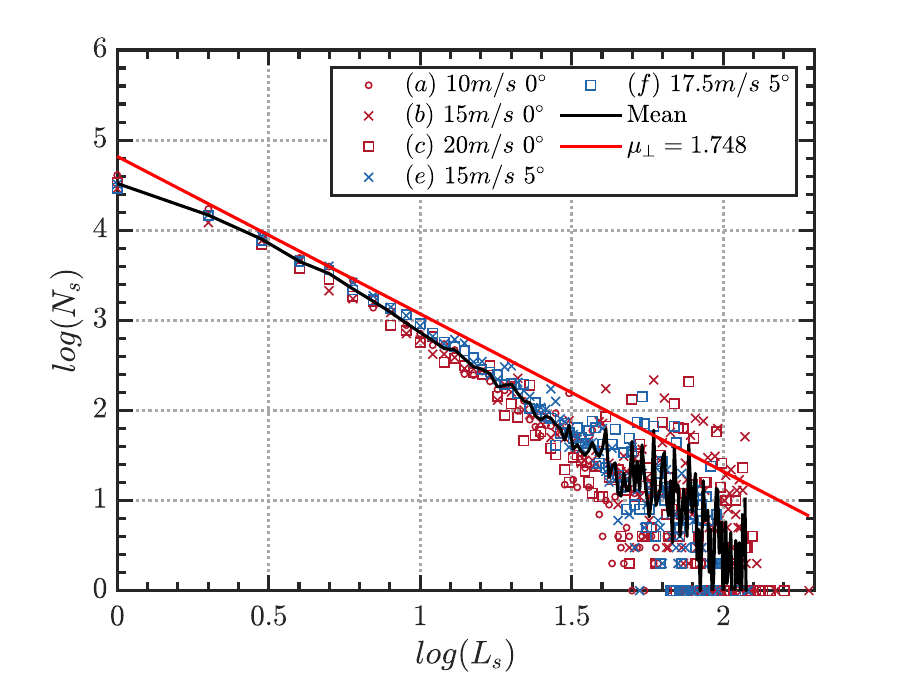}
		\caption{Cluster size distribution for spatial cluster sizes ($\perp$). The black line represents the mean of all measurements. The red line indicates the theoretical slope for (1+1)D directed percolation.}
		\label{fig:Cluster_space}
	\end{figure}

\begin{figure}[h]
		\centering
		\includegraphics[width=3.1in]{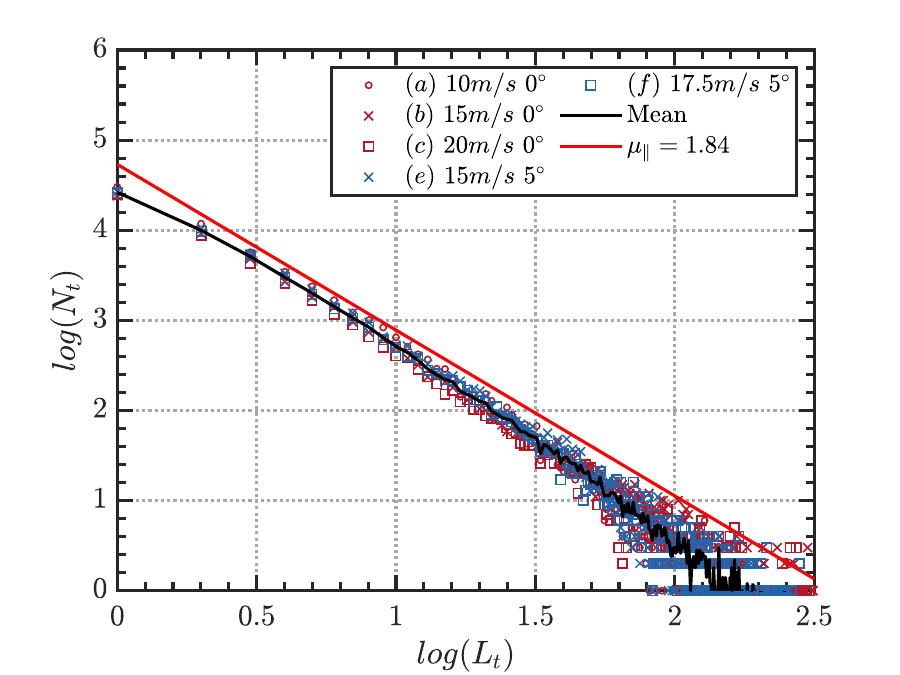}
		\caption{Cluster size distribution for temporal cluster sizes ($\parallel$) for all measurements. The black line represents the mean of all measurements. The red line indicates the theoretical slope for (1+1)D directed percolation.}
		\label{fig:Cluster_time}       % Give a unique label
	\end{figure}

	This study is based on the question whether the laminar-to-turbulent transition can be described by an universality class of directed percolation. We summarize next the central features of the percolation theory \cite{chate1987transition,Grimmett_1999,hinrichsen2000non,rupp2003critical} and link these to the specific features of the transition on an airfoil boundary layer. The focus is on the universality class of the (1+1)D percolation, as in previous studies \cite{traphan2018aerodynamics} this class was proposed as the appropriate one.\\
	In percolation theory, the system is represented by cells, which can exist in one of two states: "on" or "off." In the context of a transient boundary layers, these states correspond to turbulent (on) or laminar (off) conditions, respectively. The probability of occurrence of turbulent cells, or the turbulent fraction $\rho$, is described by an order parameter $p$ within percolation theory. In our fluid dynamical context the chord length-based Reynolds number $Re_x$ (refer to Eqn. (\ref{eqn:1})) is used as order parameter. The relationship between the turbulent fraction $\rho$ and the order parameter $Re_x$ is defined as follows:
	\begin{equation}
		\centering
		\rho(Re_x) = \rho_0 \left( \frac{Re_x-Re_c}{Re_c} \right) ^{\beta},
		\label{eqn:2}
	\end{equation} 
	like it was already used in \cite{Sano2016,traphan2018aerodynamics}. \\
	The critical Reynolds number, $Re_c$, plays a crucial role in determining the transition location between a non-percolating and a percolating system.	The parameter $\rho_0$ serves as a normalization factor, while the exponent $\beta$ is one of three universal exponents that characterize the universality class and depend on the dimensions of the system under consideration. Specifically, for (1+1)D directed percolation, the exponent $\beta$ is given as $\beta_{(1+1)D} = 0.276$ \cite{hinrichsen2000non}.\\	
	The other two  universal exponents, $\nu_{\perp}$ and $\nu_{\parallel}$, are related to the spatial distribution of cells with the same state. The exponents characterize the divergent behavior of the correlation length, $\xi_{\perp,\parallel}$, in different directions of the system at the critical point $Re_c$. The relationship describing these correlation lengths is given by:
	\begin{equation}
		\centering
		\xi_{\perp} \propto |Re-Re_c|^{\nu_{\perp}}, \ \xi_{\parallel} \propto |Re-Re_c|^{\nu_{\parallel}}.
		\label{eqn:3}
	\end{equation}
	The exponents $\nu$ for (1+1)D percolation theory have been worked out in previous research as $\nu_{\perp,(1+1)D} = 1.097$ and $\nu_{\parallel,(1+1)D} = 1.733$ \cite{hinrichsen2000non}. It is also possible to consider cluster sizes instead of correlation length scales, using the hyper scaling relation \cite{STAUFFER_1999}. The cluster sizes scale with the exponent
	\begin{equation}
		\mu_{\perp,\parallel} = 2 - \frac{\beta}{\nu_{\perp,\parallel}}.
		\label{eqn:4}
	\end{equation}  \\	
	The specific situation of the laminar-turbulent transition on an airfoil necessitates to discuss in some more details the role of the order parameter. Firstly, it is expected that $Re_c$ should align with the previously calculated transition point, denoted as $\widetilde{Re_c}$ in the previous chapter. Secondly, the order parameter varies along the chord of the airfoil, thus the cluster statistics has to be taken for different $Re_x$-positions. To get the transversal and longitudinal spatial features of the clusters we use the span wise y-direction and the temporal evolution at each $Re_x$-position.\\	
	In order to determine the states of the boundary layer (laminar (off)/turbulent (on)) from the DIT measurements, each thermogram is subjected to the binarization process using a threshold. In this process, temperatures above the threshold correspond to the laminar state (low heat transfer), while temperatures below correspond to the turbulent state (high heat transfer). The threshold value is chosen to be equal to the temperature at the previously determined inflection point of the respective measurement shown in Fig. \ref{fig:3} and Tab. \ref{tab:1}. In cases where an inflection point cannot be determined, the temperature value at the left edge of the FOV is utilized as the threshold, since the system is already transitioned. An example of a binarized thermogram is given in Fig. \ref{fig:5}.
	\\
	The turbulent fraction $\rho = \rho(Re_x,y)$ can be calculated by taking the average of all binarized DIT images, see Fig. \ref{fig:6}. The increase of the turbulent fraction from $0$ to $1$ is close to a step function and changes hardly in y-direction. 
	\\
	For further analysis, $\rho(Re_x,y)$ is averaged along the span wise direction (y-direction). The result is shown in Fig. \ref{fig:7} for the corresponding measurements of Fig. \ref{fig:3} (b),(e) and (h). The turbulent fraction indicates a very defined transition from a laminar, non-percolating system ($\rho= 0$) to a fully turbulent, percolating system ($\rho= 1$). For the case (h) $\rho$ does not start at zero since the transition takes place upstream of the FOV.
	\\
	In addition to the representation of the turbulent fraction, Fig. \ref{fig:7} also includes fits based on Eqn. (\ref{eqn:2}), depicted as solid red lines. The red symbols on each curve mark the data points used in the fitting procedure, where $\beta$ remains fixed at its theoretical value. To further illustrate the agreement with Eqn. (\ref{eqn:2}), Fig. \ref{fig:LogLogTurbFrac} shows the normalized turbulent fraction over $\epsilon= \frac{Re_x-Re_c}{Re_c}$ for all nine measurements. Notably, there is a clear clustering observed around the theoretical slope of $\beta=0.276$.
	\\
	The transition point $Re_c$ can be obtained from the fits as a fitting parameter. For all measurements, the transition points are summarized in Tab. \ref{tab:1}. The given error $\Delta Re_c$ is extracted as the confidence interval of 95\% from the fit. Comparing the calculated values of $Re_c$ we find a very good accordance to the transition points $\widetilde{Re_c}$ calculated directly from the DIT measurements.\\
	As another quantity the distribution of cluster sizes is investigated with respect to the universal exponents. Since the relations from Eqn. (\ref{eqn:3}) are only valid at the critical point, the cluster sizes need to be extracted at determined $Re_c$ values. If $Re_c$ is outside the investigated FOV, as is the case for the measurements of group two (Fig. \ref{fig:3} (d),(g)-(i)), the measurements are excluded from the following analysis.\\
	To determine the cluster sizes from the data, the binarized DITs are each combined into a 3D matrix, [$Re_x$, y, t], using the time as a third dimension. A cut is made through this matrix at $Re_x = Re_c$. This results in laminar and turbulent spots in a (y,t) domain as shown in Fig. \ref{fig:Cluster_extraction}(b). Furthermore from this presentation the cluster sizes  in y-direction (spatial, $L_s$) and t-direction (temporal, $L_t$) can be determined. The figure also shows the distribution of laminar and turbulent cells for the cases $Re<Re_c$ (Fig. \ref{fig:Cluster_extraction}(a)) and $Re>Re_c$ (Fig. \ref{fig:Cluster_extraction}(c)).
	\\
	The number $N_s(L_s)$ of clusters of a given size $L_s$ (in the spatial direction) is shown in a double logarithmic presentation in Fig. \ref{fig:Cluster_space}.  The theoretically expected slope of $\mu_\perp =1.748$ is given by a red line and shows a good agreement, which is particularly evident when considering the mean value of all cluster distributions, depicted as a black line. The presence of noise, particularly for large clusters, is attributed to the infrequent occurrence of such large structures which causes an higher statistical uncertainty.
	\\
	The scaling of the temporal cluster size distributions in the parallel direction are depicted in Fig. \ref{fig:Cluster_time}. These distributions are compared with the theoretically predicted scaling with $\mu_{\parallel} = 1.84$, expected for (1+1)D percolation.  Similar to the spatial cluster size distributions of Fig. \ref{fig:Cluster_space}, the temporal cluster size distributions show a strong alignment with the theoretically anticipated value, remarkably, even for larger cluster sizes. This improvement can be attributed to the significantly greater extension of the system in the temporal direction, as the spatial direction is limited by the FOV, whereas the temporal direction by the time of measurement.\\	
	It is worth noting that this power law behavior of cluster sizes will exhibit sensitive changes as the chosen $Re_x$ deviates from $Re_c$. Thus assuming the validity of the (1+1)D percolation and their exponents, the expected power law behaviors can also be used to determine $Re_c$. 
	\\		
	
	\section{\label{chap:Summary}Summary and Conclusion}
	This study explores the laminar-turbulent transition in the near-surface flow of an airfoil across a wide range of velocities ($u_\infty$) and angles of attack ($\alpha$). The main focus is on the question how far the directed percolation is a suitable model to describe this transition. These investigations lead to the following  new results:
	
	\begin{itemize}
		\item The experimental approach of differential image thermography (DIT) serves as the method of choice for assessing  the boundary layer's state. DIT enables to measure highly resolved in space and time the dynamic heat transfer properties inherent to various flow conditions. DIT enables measurements along curved surfaces. Overall it is this method which has lead to the quality of new results.
		
		\item All characteristic features of the presented experimental results are well in accordance with (1+1)D directed percolation. This is shown on the one hand by the fact that all results are compatible with (1+1)D directed percolation, and on the other hand, if we use our data to determine the critical exponents from fits, we get the following values:
		\begin{center}
		\begin{tabular}{|c|c|}
			\hline
			Exponent & Value from Experiment \\
			\hline
			$\beta$ & $0.2732 \pm 0.0032$ \\
			\hline
			$\mu_\perp$ & $-1.75 \pm 0.13$ \\
			\hline
			$\mu_\parallel$ & $-1.853 \pm 0.067$ \\
			\hline
			$\nu_\perp$ & $-1.09 \pm 0.58$  \\
			\hline
			$\nu_\parallel$ & $-1.85 \pm 0.87$ \\
			\hline
		\end{tabular}
	\end{center}
		Here it should be highlighted that there is no experimental study so far that has been able to show the agreement of measured data with the exponent $\mu_{\parallel}$ over such a large range. From these results higher dimensional directed percolation like (2+1)D can be ruled out. 
		
		\item For different angles of attack and flow speeds we did not find significant changes of the characterizing exponents, which confirms the idea of directed percolation theory that there are universality classes. One plausible explanation for the robustness of the (1+1)D directed percolation may emerge from the presence of pressure gradients and the resulting flow acceleration along the surface of the airfoil. This leads to stretching and subsequently flattening of the boundary layer flow. Changing experimental conditions results in a deplacement of the transition but not in a change of its nature.
		
		\item The critical Reynolds numbers of the transition determined by two methods coincide well. The methods are based on the curvature of the time-averaged differential temperatures (or DIT) and on an critical order parameter of the percolation theory. 
		The relationship between the turbulent fraction and order parameter described by the universal exponent $\beta$ is further used to calculate the critical Reynolds number $Re_c$. $Re_c$ corresponds to the transition point and can be determined much more precisely by this evaluation compared to the DIT. The relative accuracy is improved by more than one order of magnitude (see Tab. \ref{tab:1}). Furthermore, this theoretical relationship can be used to determine transition points outside the FOV.
		
		\item Directed percolation is commonly described by global order parameter changing the system from an unconnected to a percolating state. For the airfoil experiments the order parameter changes along the chord. In this sense the system may be considered as instationary, as fluid elements follow a path in flow direction with changing order parameter $Re_x$. In spite of this instationarity for a fixed location, i.e. order parameter $Re_x$ we find quasi-stationary features of the directed percolation. 
		
	\end{itemize}
	As a final remark, we want to point out that although we have found quite remarkable accordance of our data with predictions of the the directed percolation theory, we did not show that there is an underlying sub-critical bifurcation for this laminar-turbulent transition. So far one can only conclude, that the directed percolation theory reflects many experimental features. Thus it seems 
	to be a simple promising model for the highly dynamic and nonlinear processes in the boundary layer close to the laminar-turbulent transition. It could be easily integrated into CFD models to simulate the transition more realistically. A significant advantage appears to be the robustness of the model,  emphasized by the fact that the exponents agree for a wide range of experimental conditions, having potentially far-reaching benefits for industrial applications (e.g., wind turbine modeling). It is advisable to conduct more detailed investigations in future studies, focusing on the angle of attack $\alpha$ and the velocity $u_\infty$. From these, parameter spaces could be extracted that could serve as a lookup table for even more detailed descriptions.
		
	\section*{Acknowledgements}
	The authors would like to expressly thank Enno Bösenberg, who was substantially helping during in the measurements. \\
	We would also like to thank Pedro Lind and Aura Daniela Moreno Mora for fruitful and helpful discussions during data analysis and interpretation. \\
	In addition we acknowledge the Lower Saxony Ministry of Science and Culture (MWK) for financial support of this study.\\
	
	\bibliography{Percolation_Thermography.bib}
	
	\section*{Author contributions}
	All authors contributed to the study conception and design. Material preparation, data collection and analysis were performed by Tom T. B. Wester. Joachim Peinke and Gerd G\"{u}lker have contributed significantly to the development of the results through discussions. The first draft of the manuscript was written by Tom T. B. Wester and all authors commented on previous versions of the manuscript. All authors read and approved the final manuscript.
	
	\section*{Competing interests}
	The authors declare no competing interests.
	
\end{document}